\documentclass[twoside]{article}
\usepackage{qic,epsfig}

\begin{document}
\setlength{\textheight}{8.0truein}    

\runninghead{Universal quantum computation with quantum-dot cellular
automata in decoherence-free subspace}
            {Z.-Y.Xu, M.Feng, and W.-M.Zhang}

\normalsize\textlineskip
\thispagestyle{empty}
\setcounter{page}{1}

\copyrightheading{0}{0}{2008}{000--000}

\vspace*{0.88truein}

\alphfootnote

\fpage{1}

\centerline{\bf
UNIVERSAL QUANTUM COMPUTATION WITH QUANTUM-DOT CELLULAR}
\vspace*{0.035truein} \centerline{\bf AUTOMATA IN DECOHERENCE-FREE
SUBSPACE} \vspace*{0.37truein} \centerline{\footnotesize
Z. Y. XU} \vspace*{0.015truein} \centerline{\footnotesize\it State
Key Laboratory of Magnetic Resonance and Atomic and Molecular
Physics, Wuhan Institute of Physics} \vspace*{0.015truein}
\centerline{\footnotesize\it and Mathematics, Chinese Academy of
Sciences, Wuhan, 430071, China} \vspace*{0.015truein}
\centerline{\footnotesize\it Graduate School of the Chinese Academy
of Sciences, Beijing, 100049, China} \vspace*{10pt}
\centerline{\footnotesize M. FENG\footnote{mangfeng@wipm.ac.cn}}
\vspace*{0.015truein} \centerline{\footnotesize\it State Key
Laboratory of Magnetic Resonance and Atomic and Molecular Physics,
Wuhan Institute of Physics} \vspace*{0.015truein}
\centerline{\footnotesize\it and Mathematics, Chinese Academy of
Sciences, Wuhan, 430071, China} \vspace*{10pt}
\centerline{\footnotesize W. M. ZHANG} \vspace*{0.015truein}
\centerline{\footnotesize\it Department of Physics and Center for
Quantum Information Science, National Cheng Kung University,}
\vspace*{0.015truein} \centerline{\footnotesize\it Tainan 70101,
China} \vspace*{0.225truein} \publisher{(received date)}{(revised
date)}

\vspace*{0.21truein}

\abstracts{
We investigate the possibility to have electron-pairs in
decoherence-free subspace (DFS), by means of the quantum-dot
cellular automata (QCA) and single-spin rotations, to carry out a
high-fidelity and deterministic universal quantum computation. We
show that our QCA device with electrons tunneling in two dimensions
is very suitable for DFS encoding, and argue that our design favors
a scalable quantum computation robust to collective dephasing
errors.}{}{}

\vspace*{10pt}

 \vspace*{3pt}
\communicate{to be filled by the Editorial}

\vspace*{1pt}\textlineskip    

Spin degrees of freedom of electrons in quantum dots have been
considered as good candidates to encode qubits over past years, due
to their long decoherence time and full controllability. Of
particular interest is the recent achievements of ultrafast
manipulation of electron spin in conduction band of the quantum dot
\cite{gupta, kroutvar} and coherent tunneling of electrons between
neighboring quantum dots \cite{haya,petta}. These technical
progresses have led to more and more concerns on movable electron
based quantum gates.

It has been shown in \cite{been,engle,zhang1,zhang2} that, by
manipulating spin degrees of freedom, we are able to perform
universal quantum computation (UQC) with movable electrons. The key
idea is that we encode qubits in the electron spins, but make
measurement \cite{been,engle} or make entanglement
\cite{zhang1,zhang2} by means of the electron charges. As spin and
charge (i.e., orbit) degrees of freedom commute, a measurement on
the charge of an electron does not affect the spin of the electron.
With these ideas, we could use the movable (or say, free) electrons
to entangle the spin states of different electrons
\cite{been,engle,zhang1,zhang2,feng1}, to analyze the multipartite
entanglement \cite{been,feng1}, and to purify the existing
entanglement \cite{oh}.

However, the electron spins in quantum dots severely suffer from
decoherence regarding environmental noise, such as the surrounding
nuclear spins, background charge fluctuation and noise,
electron--phonon interaction, low-frequency noise and so on
\cite{explain} and it is evident that the confinement of the quantum
dots makes the decoherence enhanced. Fortunately, as long as our
operation on the electron spin is quick enough, the influence from
the surrounding nuclear spins, also called Overhauser effect, could
be effectively considered as from a constant magnetic field
\cite{ima0}. In principle, we may employ spin echo technique to
remove any errors due to constant magnetic field, whereas
experimental evidence has shown that the spin echo pulses could not
fully eliminate errors regarding dephasing \cite{petta}. To defeat
decoherence, people have worked out a number of ideas, such as in
\cite{palma,duan,lidar1,kempe,lidar2,hwang,lidar3}, where the error
avoiding strategies carried out in decoherence-free subspaces (DFS)
are useful for suppressing collective dephasing errors, and are
relatively simpler, because they only require some special encodings
immune from certain system-environment disturbances but no error
correction steps are needed.

We focus in the present work on the recent proposal with quantum-dot
cellular automata (QCA), in which spin entanglement of different
electrons could be achieved without spin-spin interaction
\cite{zhang1,zhang2}. We will show that UQC in DFS could be carried
out by QCA settings in a relatively simpler way than by other
systems. QCA was originally proposed as a transistorless alternative
to digital circuit devices at the nanoscale \cite{orlov}, and then
employed in quantum systems \cite{Snider,Benjamin,Lloyd}. In the
present paper, we show our idea of DFS encoding with QCA applied to
quantum dots, in which electron spins would be involved to encode
qubits, and QCA behaves quantum mechanically with two electrons
tunneling coherently between two antipodal sites on the QCA due to
Coulomb repulsion. As shown in \cite{zhang1,zhang2}, different from
the free-electron QC models under screening assumption
\cite{been,engle}, the QCA quantum computation makes use of the
Coulomb interaction between electrons throughout the operations, and
could thereby achieve deterministic entanglement between electron
spins.

The DFS we employ is spanned by the encoding states
$|0_{L}\rangle=|01\rangle$ and $|1_{L}\rangle=|10\rangle$ with
$|0\rangle$ and $|1\rangle$ the spin up and down states of the
electron in the dot, respectively, which constitutes a well-known
DFS scheme immune from dephasing induced by the system-environment
interaction in the form of $Z\otimes B$, where
$Z=\sigma_{z}^{1}\oplus \sigma_{z}^{2}$ and B is a random bath
operator. For clarity, we will call $|0_{L}\rangle$
($|1_{L}\rangle$) logic qubit and $|0\rangle$ ($|1\rangle$) physical
qubit. As there is no spin-spin coupling between the electrons, we
have degeneracy between $|0_{L}\rangle$ and $|1_{L}\rangle,$
implying that no noise from collective dephasing would affect the
encoded subspace we employ. It also means that the dot-dot spacing
in our design must be bigger than those in
\cite{petta,klauser,romito}. As collective errors due to coupling to
environment are generally considered to be the main problem in
solid-state system at low temperature \cite{lidar4}, we will focus
throughout this work on overcoming the collective dephasing. Other
noise beyond collective dephasing could also be removed by some
additional operations, as shown later. We will demonstrate that
three basic logic gates for a UQC could be carried out by our
DFS-encoded electron spins, without any auxiliary spin qubit
required. As dephasing is strongly suppressed and the QC is run
strictly within the DFS, the entangled state generated in our design
could be kept in high fidelity for a long time.

\begin{figure} [htbp]
\centerline{\epsfig{file=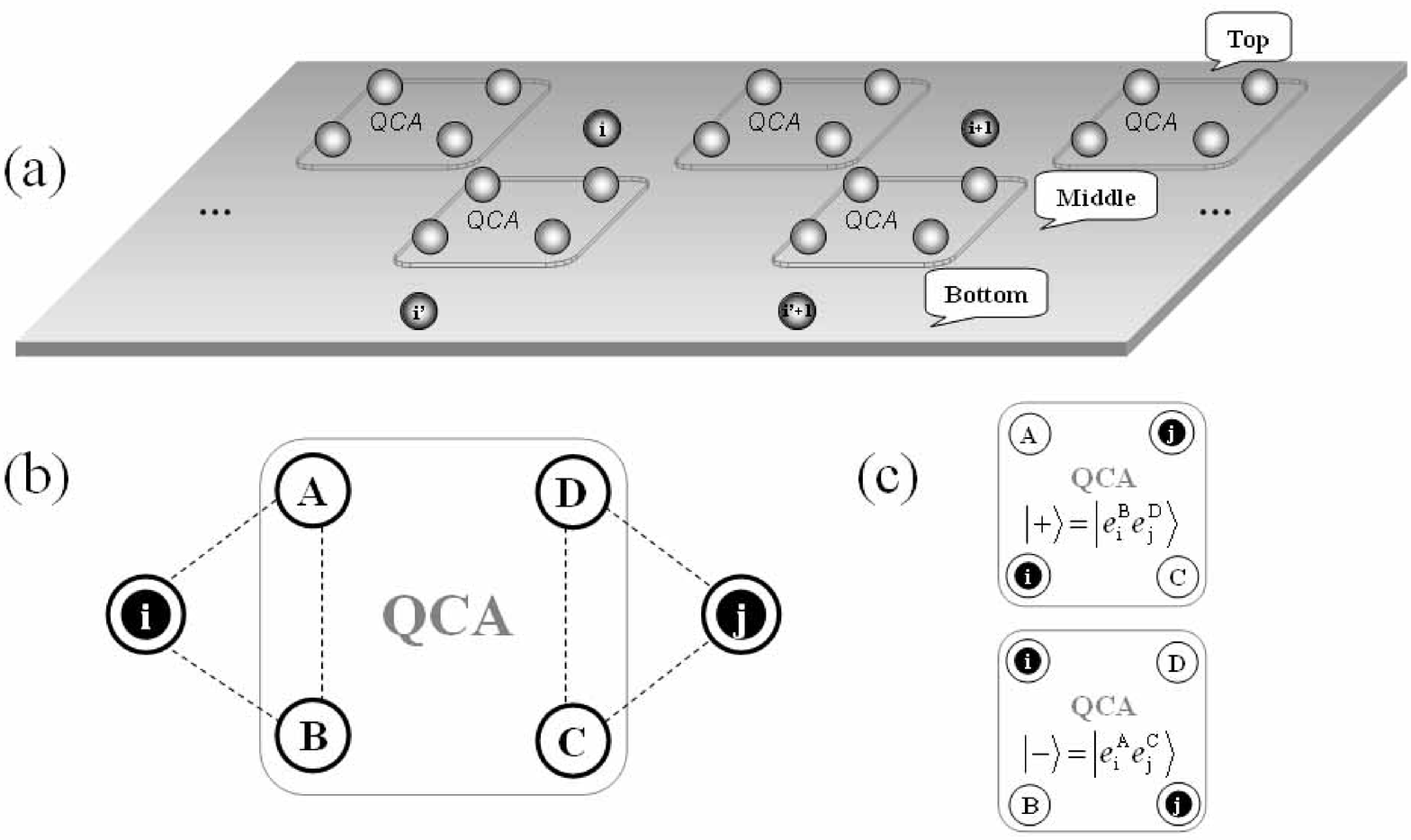, width=9.2cm}} 
\vspace*{13pt} \fcaption{\label{motion}(a) Schematics for our
proposed design, where the dots i, with
$i=1,2,3,....$, are initially prepared in $|0\rangle,$ and dots \textit{i}%
$^{^{\prime}}$, with
$i^{\prime}=1^{^{\prime}},2^{^{\prime}},3^{\prime},....,$ are
initially in $|1\rangle.$ The logic qubits are constructed by
$|\rangle_{ii^{\prime}}.$ (b) Four initially neutral quantum dots in
the square constitute a QCA, and two initially charged quantum dots
are separated by the QCA, where the black dots represent single
electrons and the dashed lines connecting quantum dots denote
possible tunnelings. (c) Coulomb
repulsion causes two full polarized charge states $|+\rangle=|e_{i}^{B}%
e_{j}^{D}\rangle$ and $|-\rangle=|e_{i}^{A}e_{j}^{C}\rangle$, where
A, B, C and D mean the sites and \textit{i} and \textit{j} denote
the different electron spins.}
\end{figure}

As shown in Fig.~1(a), the QCA blocks and the quantum dots encoding
the qubits are arranged in alternate way in two dimensions, where
the large spacing between the qubits, e.g., hundreds of nanometers
or even of the order of micrometer to make the spin-spin interaction
negligible, is helpful for individual manipulation on the qubits. To
have a UQC in DFS, we have to construct three logic-qubit gates. The
first is the Hadamard gate $H_{L}:$
$|0_{L}\rangle=|01\rangle_{ii^{\prime}}\Rightarrow\frac{1}{\sqrt{2}}%
(|0_{L}\rangle+|1_{L}\rangle)=\frac{1}{\sqrt{2}}(|01\rangle_{ii^{\prime}%
}+|10\rangle_{ii^{\prime}}),$ and $|1_{L}\rangle=|10\rangle_{ii^{\prime}%
}\Rightarrow\frac{1}{\sqrt{2}}(|0_{L}\rangle-|1_{L}\rangle)=\frac{1}{\sqrt{2}%
}(|01\rangle_{ii^{\prime}}-|10\rangle_{ii^{\prime}}),$ where $i$ and
$i\prime$ denote different dots with $i,i\prime=1,2,3,....$ The
second gate is the two-logic-qubit conditional gate. We will
construct a controlled-phase flip (CPF) as an example, i.e., a phase
$\pi$ appearing as the prefactor of $|1_{L}1_{L}\rangle$ after the
gating, and the third one is for a single-logic-qubit rotation
$Q_{L}(\theta),$ i.e., $a|0_{L}\rangle +b|1_{L}\rangle\Rightarrow
a|0_{L}\rangle+be^{i\theta}|1_{L}\rangle.$

Consider the initial state of the two electrons in quantum dots $i$
and $j$ to be $|e_{i}e_{j}\rangle\otimes|S_{i}S_{j}\rangle,$ where
$|e_{i}e_{j}\rangle$ are charge states to be auxiliary,
$|S_{i}S_{j}\rangle$ are spin states for qubit encoding, and $j$
could be $i^{\prime}$ in the case of $H_{L}$ gating or $i+1$ for
achieving CPF. By the same operations as in \cite{zhang2}, after the
electrons tunnel to dots A and C, we switch off the channels between
the dots i, j and the QCA, and then turn on the bias for the
electron tunneling between the sites A and B, and between the sites
C and D (See Fig. 1(b)). We may describe the quantum behavior on the
QCA by following Hamiltonian in units of $\hbar=1$
\cite{haya,zhang2},
\begin{equation}
H_{QCA}=\frac{\omega_{0}}{2}(|+\rangle\langle+|-|-\rangle\langle
-|)+\frac{\gamma}{2}(|+\rangle\langle-|+|-\rangle\langle+|),
\end{equation}
where $|+\rangle=|e_{i}^{B}e_{j}^{D}\rangle$ and $|-\rangle=|e_{i}^{A}%
e_{j}^{C}\rangle$ are polarized charge states defined in
\cite{zhang1,zhang2} and in Fig. 1(c). $\omega_{0}$ represents the
energy offset of the polarized states $|\pm\rangle$ from the balance
of on-site potential, Coulomb repulsion and the external bias
energy. $\gamma$ accounts for the tunneling between these two
polarized states.

To carry out the first gate $H_{L}$, we set $\omega_{0}$ to be zero
(i.e., a symmetric QCA) and start the tunneling from the state
$|-\rangle\otimes |S_{i}S_{i^{^{\prime}}}\rangle$ where the
subscripts correspond to the dots the electrons come from, and the
electron with spin $|S_{i}\rangle (|S_{i^{^{\prime}}}\rangle)$ will
tunnel between A(C) and B(D). During the electron tunneling on the
QCA, we perform single-spin rotations U$_{BD}$ and U$_{AC}$ (defined
later) on the electronic states at the sites B, D and A, C. As the
tunneling is coherent, these single-spin operations could be done
simultaneously \cite{zhang2}. At $t=\pi/2\gamma,$ we stop our
operations on the QCA, and drive the electrons back to dots $i$ and
$i^{^{\prime}}$
\cite{zhang2}$.$ Then we get%
\[
|e_{i}e_{i^{^{\prime}}}\rangle\otimes\frac{1}{\sqrt{2}}(U_{AC}-iU_{BD}%
)|S_{i}S_{i^{^{\prime}}}\rangle,
\]
where $U_{AC}=R_{x}^{A}(\pi)\otimes R_{x}^{C}(3\pi)$ and $U_{BD}=R_{z}%
^{B}(3\pi)\otimes I^{D},$ with the superscripts being the sites
where the electron is rotated,
$R_{k}(\theta)=\exp(-i\theta\sigma_{k}/2),$ $k=x,y,z$, and $I$ being
an identity operator$.$ As electrons tunnel back from dots A(C) and
B(D) to $i(i\prime)$ simultaneously \cite{zhang2}, we may simply
rewrite
$U_{AC}$ and $U_{BD}$ as $U_{AC}=R_{x}^{i}(\pi)\otimes R_{x}^{i^{\prime}}%
(3\pi)$ and $U_{BD}=R_{z}^{i}(3\pi)\otimes I^{i^{\prime}}.$ It is
easy to
verify that above operations yield $H_{L}:a|0_{L}\rangle+b|1_{L}%
\rangle\Rightarrow$
$\frac{1}{\sqrt{2}}[a(|0_{L}\rangle+|1_{L}\rangle)$
$+b(|0_{L}\rangle-|1_{L}\rangle)],$ as shown in Fig. 2(a).

\begin{figure} [htbp]
\centerline{\epsfig{file=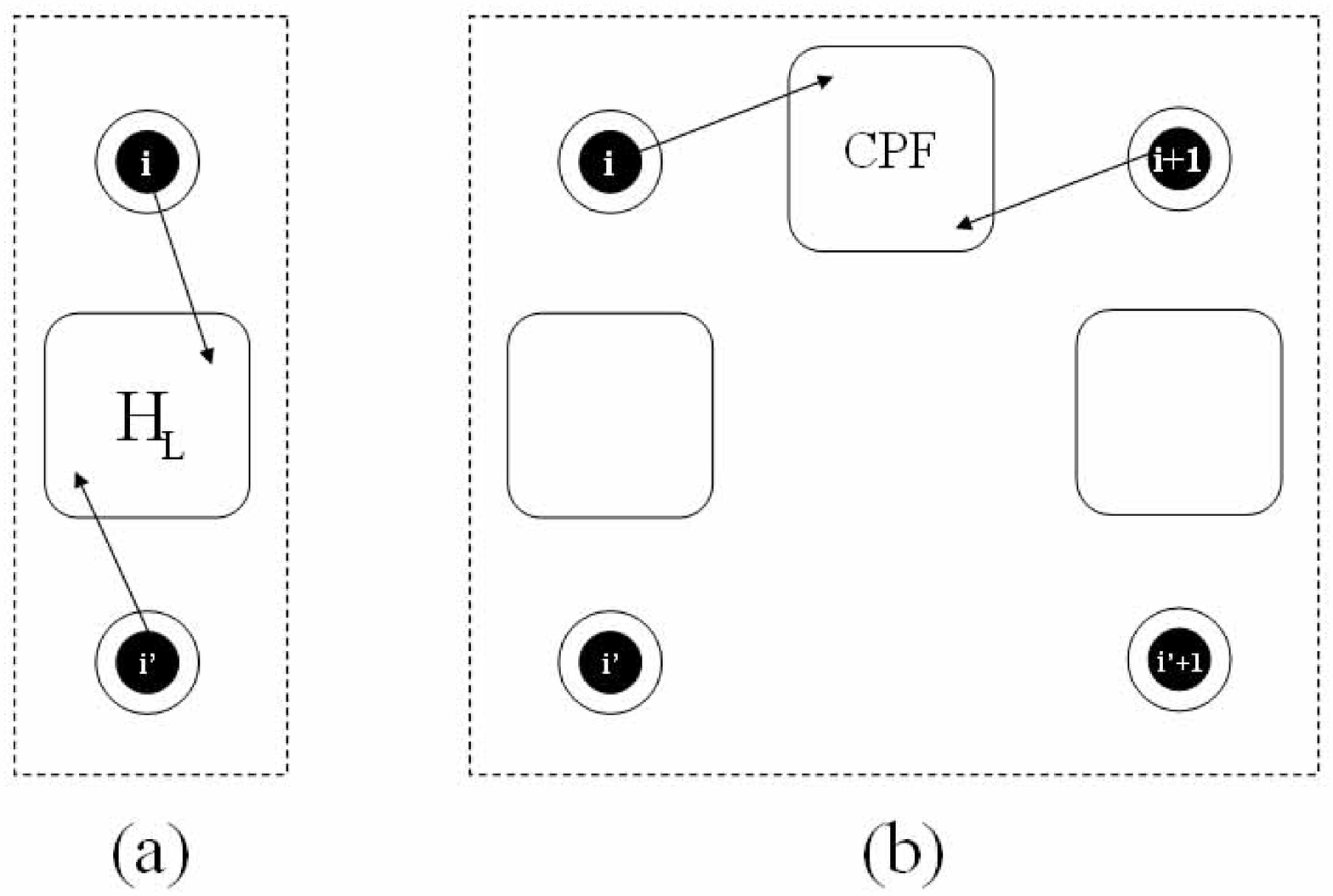, width=8.5cm}} 
\vspace*{13pt} \fcaption{\label{motion}Logic-qubit quantum gates,
where (a) is for $H_{L}$ carried out between dots $i$ and
$i^{\prime}$. (b) is for CPF between dots $i$ and $i+1$ in the top
line.}
\end{figure}

The second gate happens between pairs $i-i^{\prime}$ and
$(i+1)-(i^{\prime }+1)$ of the initial state
$(a|01\rangle_{i,i^{\prime}}+b|10\rangle _{i,i^{\prime}})\otimes$
$(c|01\rangle_{i+1,i^{\prime}+1}$ $+d|10\rangle
_{i+1,i^{\prime}+1}).$ The CPF yields%
\[
ac|0101\rangle_{i,i^{\prime},i+1,i^{\prime}+1}+ad|0110\rangle_{i,i^{\prime
},i+1,i^{\prime}+1}+bc|1001\rangle_{i,i^{\prime},i+1,i^{\prime}+1}%
-bd|1010\rangle_{i,i^{\prime},i+1,i^{\prime}+1},
\]
which is actually equivalent to a CPF on electrons $i$ and $i+1$ in
the top line (See Fig. 2(b)). To achieve such a CPF, we may employ
the controlled-NOT (CNOT) gate in \cite{zhang2} sandwiched by two
Hadamard gates on the target physical qubit. But we hope to
accomplish the CPF directly to make our implementation simple. So
under the Hamiltonian $H_{QCA}$ with $\omega_{0}=0$ and the initial
state $|-\rangle\otimes|S_{i}S_{i+1}\rangle,$ we start the tunneling
assisted with single-spin rotations$.$ Like in above H$_{L}$ gating,
we stop the electron tunneling at $t=\pi/2\gamma,$ and drive the
electrons back to the dots $i$ and $i+1$. So we have
$|e_{i}e_{i+1}\rangle\otimes
\frac{1}{\sqrt{2}}(\bar{U}_{AC}-i\bar{U}_{BD})|S_{i}S_{i+1}\rangle,$
with $\bar{U}_{AC}=R_{z}^{A}(\pi/2)\otimes
R_{z}^{C}(\pi/2)=R_{z}^{i}(\pi/2)\otimes
R_{z}^{i+1}(\pi/2)$ and $\bar{U}_{BD}=R_{z}^{B}(3\pi/2)\otimes R_{z}^{D}%
(3\pi/2)=R_{z}^{i}(3\pi/2)\otimes R_{z}^{i+1}(3\pi/2),$ which yields%
\begin{equation}
|\varphi\rangle=\frac{1}{\sqrt{2}}|e_{i}e_{i+1}\rangle\otimes(1-i)\left(
\begin{array}
[c]{cccc}%
1 & 0 & 0 & 0\\
0 & 1 & 0 & 0\\
0 & 0 & 1 & 0\\
0 & 0 & 0 & -1
\end{array}
\right)  |S_{i}S_{i+1}\rangle.
\end{equation}
If we neglect the additional global phase, we have fulfilled the CPF
operation: $|S_{i}S_{i+1}\rangle\Rightarrow(-1)^{S_{i}S_{i+1}}|S_{i}%
S_{i+1}\rangle$ between the dots $i$ and $i+1,$ with $S_{i}$,
$S_{i+1}=0,1.$ This physical-qubit CPF also implies the logic-qubit
CPF between pairs $i-i^{\prime}$ and $(i+1)-(i^{\prime}+1),$ as
shown in Fig. 2(b).

The third gate $Q_{L}(\theta)$ could be achieved by rotating one of
the physical qubits. So different from the first and the second
gates, the implementation of the third gate employs Faraday rotation
\cite{meier,ima0}, instead of the tunneling on QCA. We apply
$\sigma_{(z)}^{+}$ polarized light on a certain dot in the bottom
line. A phase $e^{i\delta_{0}}$ ($e^{i\delta _{1}})$ will be created
if the electron spin of the dot is initially $|0\rangle$
($|1\rangle$), due to virtual excitation of exciton including heavy
(light) hole state \cite{meier,ima0}. As $\delta_{0}$ is larger than
$\delta_{1}$ and both of them could be exactly controlled, we could
achieve $Q_{L}(\theta)$ with $\theta=\delta_{0}-\delta_{1}.$

With the three basic gates above, we could carry out a universal
quantum gating with the electron pairs. However, in terms of
DiVincenzo's checklist \cite{divi}, a UQC also requires high-quality
preparation of initial states and the efficient readout, besides the
universal quantum gating. In our case, the initial qubit states on
the top line should be in $|00...0\rangle$ (i.e., all spins up) and
the qubit states on the bottom line are initially $|11...1\rangle$
(i.e., all spins down), which correspond to the logic state
$|0_{L}\rangle.$ As the interdot separation is big, this job could
be accomplished individually by the techniques in
\cite{kroutvar,finley}, where a single conduction band electron was
produced \cite{finley} and single-spin manipulation on the
conduction band electron has been achieved \cite{kroutvar}. The
single-spin rotation could also be made by ultrafast laser pulses
which accomplish substantial and accurate spin rotation at the
timescale of femtosecond \cite{gupta}. The efficient readout of
qubit states has been available optically by nondestructive
detection of the electron spin in the conduction band of the quantum
dot \cite{bere}. The same job could also be done by single-shot
technique \cite{elzer} based on the charge signal due to electron
jumping. If the electron could jump back to the original site after
the detection, this readout is also nondestructive \cite{zhang2}.
Therefore, up to now, we have proved that a UQC with the DFS encoded
electron pairs is available in our QCA-based device.

As it strongly suppresses the collective dephasing, the DFS encoding
could much reduce the operations for spin echo and thereby actually
reduce the gating time and enhance the fidelity, although it seems
to increase the overhead resource. Besides collective dephasing,
however, there would be other dephasing errors in a real system,
such as logic errors regarding $\sigma_{r}^{i}\sigma_{r}^{i+1}$ with
$r=x,y,z,$ and the leakage errors
related to following undesired operations: $\sigma_{x}^{i}$, $\sigma_{x}%
^{i+1}$, $\sigma_{y}^{i}$, $\ \sigma_{y}^{i+1}$,
$\sigma_{x}^{i}\sigma
_{z}^{i+1}$, $\sigma_{z}^{i}\sigma_{x}^{i+1}$, $\sigma_{y}^{i}\sigma_{z}%
^{i+1}$, $\sigma_{z}^{i}\sigma_{y}^{i+1}$ \cite{lidar4,lidar5}. To
eliminate the logic errors, we may employ deliberately designed
Bang-Bang
pulse sequences including $\sigma_{x}^{i}\sigma_{x}^{i+1},$ $\sigma_{y}%
^{i}\sigma_{y}^{i+1},$ and $\sigma_{z}^{i}\sigma_{z}^{i+1}$
respectively, with further amendment by refocusing on individual
physical-qubits \cite{lidar4}. Bang-Bang pulses are strong and fast,
which could effectively average out the environment-induced noise,
and keep the quantum state from dissipation by repeatedly kicking
the qubit \cite{viola,kho}. The leakage errors could be in principle
fully removed by the leakage-elimination operator introduced by
\cite{lidar5}, which is actually associated with projection
operations and could be easily applied to our design. So after the
treatment above, dephasing errors could be much suppressed in our
scheme, and thereby $T_{2}$ in our design should be in principle
much longer than that without using DFS. Moreover, since the errors
are brought about by unpredictable factors, e.g., the fluctuation of
the magnetic field, we have to first use interrogative Bang-Bang
pulses to determine the required values for correction, which has
been actually a sophisticated technique \cite{vende}. A very recent
experiment \cite{briggs} has shown the power of Bang-Bang pulses to
decouple the qubit from the environmental noise. So, with the
encoding plus Bang-Bang pulses assisted sometimes by individual
physical-qubit refocusing,\ all dephasing errors could be strongly
suppressed in our design.

For other sources of decoherence beyond dephasing, the mechanism is
very complicated and not fully clear yet. They yield decoherence
time $T_{1}.$ It was reported that $T_{1}$ could be of the order of
millisec in GaAs and In(Ga)As quantum dots \cite{fuji}, and in a
preliminary experiment for QCA with two electrons involving no spin
\cite{gard}, the coherent tunneling of the electrons diminished very
quickly. Although we have not yet fully understood these decoherence
sources, lower temperature is helpful for suppressing most of them.
We have also noticed that elaborately controlled spin-echo pulses
could extend $T_{2}$ to 1 microsec \cite{petta}, and thereby we
guess that $T_{1}$ should be longer. As there is no fluctuation
regarding spin-spin exchange energy and hybridized states
\cite{romito} in our design due to negligible inter-spin coupling,
we may expect $T_{1}$ in our design to be longer than microsecond in
the low temperature.

In addition, spin-orbit interaction could also affect the spin
coherence and spin-spin coupling \cite {Khaetskii}. But in our case,
there is no direct spin-spin coupling, and the electron staying in
the ground state of the conduction band is of the $s$-orbit
wavefunction. As a result, the effect due to spin-orbit interaction
is negligible in our scheme.

There have been some preliminary QCA experiments without involving
electron spins \cite{orlov,Snider,gard}. But as the electron
tunneling involving spin degrees of freedom has been available, we
believe the QCA experiment relevant to electron spin would also be
achievable soon. In what follows, let us briefly discuss the
experimental feasibility of our proposed scheme. Using the values in
\cite{feng2}, we may assess an entangled state between the electrons
$i$ and $j$ to be achievable within 70 picosec, provided that the
electron tunneling rate on the QCA could be as fast as 200 GHz
\cite{feng2}.\ As the implementation time is much shorter than
$T_{1}$ and $T_{2},$ we may neglect decoherence in our discussion.
But due to the rapid operation, we have to pay attention to the
possible imprecision in the single-spin rotation and in the bias
voltage control. For an estimate, we have assumed in our numerical
calculation a laser induced phase error $\epsilon$ for every $\pi/2$
single-spin rotation and a phase error $\delta$ by voltage control
in each tunneling on the QCA. Fig. 3 demonstrates the fidelity of
$H_{L}$ and CPF on different states under these errors. We could
find that the error $\epsilon$ is more destructive than $\delta$,
which implies the accurate manipulation by laser to be more
essential to our implementation. Another point is that the CPF works
better than $H_{L}$ under the same condition. The reason is that
$H_{L}$ involves larger rotations which bring about more phase
errors regarding $\epsilon.$ The results remind us to pay more
attention to the operations by the ultrafast laser pulses.

\begin{figure} [htbp]
\centerline{\epsfig{file=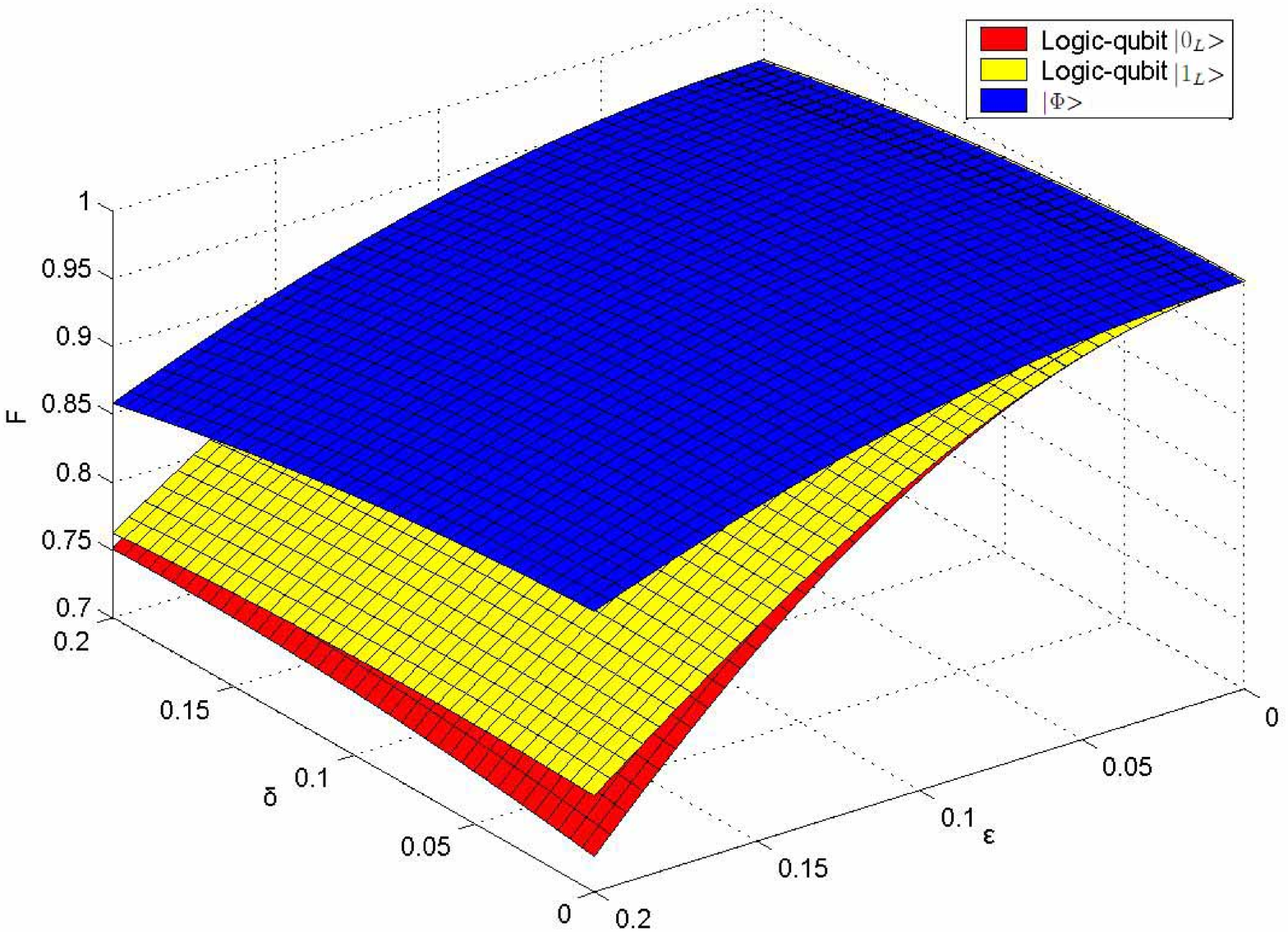, width=10.2cm}} 
\vspace*{13pt} \fcaption{\label{motion}(Color online) Numerical
simulation for the fidelity of the gating $H_{L}$ and CPF under
imprecise operations, where $\epsilon$ and $\delta$ are phase errors
induced, respectively, in the laser manipulation and the voltage
control. The curved surfaces in red and yellow represent $H_{L}$ on
logic-qubits $|0_{L}\rangle$ and $|1_{L}\rangle,$ respectively, and
the blue surface is for CPF on the state
$|\Phi\rangle=(|00\rangle+|01\rangle+$ $|10\rangle+|11\rangle)/2.$}
\end{figure}

Compared with previous devices producing entanglement between free
electrons \cite{been,engle,feng1}, our QCA-based design could
achieve the logic-qubit quantum gates straightforwardly in a simpler
fashion. For example, the CPF gating could be made with much reduced
steps compared to \cite{been,engle,feng1}. This is because that the
free electrons under screening model \cite{been,engle,feng1}
interact only by measurement, which is probabilistic, while our
implementation, under Coulomb interaction, is made straightforwardly
and deterministically. In addition, the measurement in
\cite{been,engle,feng1} is made by the time-resolved charge detector
which is technically unavailable at present, whereas no charge
detector is required in our design. Furthermore, as the movable
electrons are distinguishable throughout our scheme, no concern
about quantum characteristic of identical particles is needed. More
importantly, as dephasing errors are strongly suppressed, the
entangled states in our design could be kept in high-fidelity for a
longer time than in any proposal without using DFS.

To some extents, our scheme is similar to that with multizone trap
by moving ions \cite{kielp}. Both the electrons in our design and
the ultracold ions in the trap are exactly controllable, e.g., to be
static and moving under control. Besides, both the designs are
scalable, and deterministically operated. It has been shown in
\cite{kielp} that the DFS encoding could suppress the collective
dephasing errors in trapped ions separated by 5 $\sim10$ $\mu$m to
$10^{-4}.$ So it should work better in our design with the dots'
spacing being approximately 1 $\mu$m$.$ On the other hand, due to
controllable tunneling in a two-dimentional configuration, our
designed QCA setting is more favorable for UQC in DFS than ion traps
or other systems: We need no movement of qubits for a long distance
as in multizone trap \cite{kielp}, and the dephasing resisted UQC
could be achieved with no need of auxiliary qubits \cite{xue} and no
danger going beyond the DFS during operation \cite{feng3}. What is
more, the above proposals \cite{xue,feng3} are probabilistic due to
measurement involved, whereas our implementation is deterministic. \

In summary, we have\ demonstrated the possibility to carry out a DFS
encoding for UQC robust to collective dephasing errors by a
QCA-based device using electronic tunneling and single-spin
rotations. Our scheme only involves gate voltage controls of the
electron tunneling and optical manipulation of the electron spin in
quantum dots. To suppress other errors beyond collective dephasing,
we may employ some additional means, which are in principle
achievable in our design. Although some of the necessary steps are
still challenging with current experimental techniques, our proposed
design without spin--spin interaction and dephasing errors, but with
relatively large dot-dot spacing, is helpful for experimental
observation of coherence and entanglement of electron spins in
quantum dots and provides a promising way towards scalable QC with
quantum dots.

This work is partly supported by NNSF of China under Grant No.
10774163, and partly by the NFRP of China under Grant No.
2006CB921203.

\section{References}

\end{document}